\documentclass{article}
\usepackage{arxiv}
\usepackage[ruled,vlined]{algorithm2e}
\usepackage{amsmath, amssymb}
\usepackage{graphicx}
\usepackage{color,array}
\usepackage{cite}
\usepackage{amsmath,amssymb,amsfonts,bm}
\usepackage[utf8]{inputenc} 
\usepackage[T1]{fontenc}    
\usepackage{hyperref}       
\usepackage{url}            
\usepackage{booktabs}       
\usepackage{amsfonts}       
\usepackage{nicefrac}       
\usepackage{microtype}      
\usepackage{lipsum}		
\usepackage{graphicx}
\usepackage{natbib}
\usepackage{doi}
\usepackage{appendix}
\newtheorem{definition}{Definition}[section]
\newtheorem{theorem}{Theorem}[section]

\title{\textsc{LOTUS}: \textbf{L}AYER-\textbf{O}RDERED \textbf{T}EMPORALLY-\textbf{U}NIFIED \textbf{S}CHEDULES 
FOR QUANTUM APPROXIMATE OPTIMIZATION ALGORITHMS}

\author{ \href{https://orcid.org/0000-0000-0000-0000}{\includegraphics[scale=0.06]{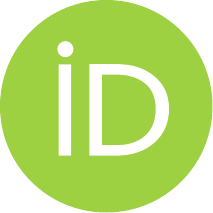}\hspace{1mm}Phuong-Nam Nguyen}\thanks{Corresponding author: namnp31@viettel.com.vn} \\
	Quantum Engineer, Viettel High Technology Industries Corporation, Viettel Group\\
	380 Lac Long Quan Street, Nhat Tan Ward, Tay Ho District, Hanoi, 100000, Vietnam. \\
	\texttt{namnp31@viettel.com.vn} 
}

\renewcommand{\shorttitle}{\scriptsize LOTUS: \textbf{L}AYER-\textbf{O}RDERED \textbf{T}EMPORALLY-\textbf{U}NIFIED \textbf{S}CHEDULES 
FOR QAOA}

\hypersetup{
pdftitle={LOTUS},
pdfsubject={q-bio.NC, q-bio.QM},
pdfauthor={Nguyen~Phuong-Nam},
pdfkeywords={First keyword, Second keyword, More},
}

\begin{document}
\maketitle

\begin{abstract}
	In this paper, we introduce LOTUS (Layer-Ordered Temporally-Unified Schedules), which is a framework that restructures QAOA from a high-dimensional, chaotic search into a low-dimensional dynamical system. By replacing independent layer-wise angles with a Hybrid Fourier-Autoregressive (HFA) mapping, LOTUS enforces global temporal coherence while maintaining local flexibility. LOTUS consistently outperforms standard optimizers, achieving up to a $27.2\%$ improvement in expectation values over L-BFGS-B and $20.8\%$ compared with COBYLA. Besides, our proposed method drastically reduces computational costs, requiring over $90\%$ fewer iterations than methods like Powell or SLSQP.
\end{abstract}


\section{Introduction}
\paragraph{Backgrounds:}
The Quantum Approximate Optimization Algorithm (QAOA) is a prominent variational quantum algorithm (VQA) designed to find approximate solutions to combinatorial optimization problems by alternating between a cost Hamiltonian $H_C$ and a mixer Hamiltonian $H_B$ \cite{farhi2014quantum,blekos2024review,choi2019tutorial,zhou2020quantum,willsch2020benchmarking,abualigah2024quantum}. While the standard QAOA ansatz treats variational angles ${\gamma_l, \beta_l}_{l=1}^p$ as $2p$ independent degrees of freedom, this discrete independence introduces significant pathologies that hinder its scalability to utility-scale quantum advantage. As the circuit depth $p$ increases, the linear growth of the parameter space triggers a dimensionality explosion. This often leads to barren plateaus \cite{mao2025qaoa,larocca2022diagnosing,rajakumar2024trainability,wang2021noise,larocca2025barren}, where gradients vanish exponentially, making it computationally expensive for traditional gradient-free and gradient-based solvers to find optimal schedules.

\paragraph{Key observations:} We contextualize the necessity for a unified parameterization by documenting the pathological behaviors observed in standard QAOA optimization landscapes (see Figure~\ref{fig:prelim_results}). Our experimental benchmarks reveal three key observations that motivate the LOTUS framework:
\begin{enumerate}
\item \textit{Discrete piecewise-constant transitions:} When optimizing the $2p$ parameters independently, the resulting schedules frequently exhibit “sudden jumps” between plateaus. For a sequence of layers $l=1$ to $p$, the optimal values of $\gamma_l$ (or $\beta_l$) tend to remain constant for several layers—forming a plateau—before abruptly shifting to a different constant value. This creates a “piecewise-constant” or step-function pattern across the circuit depth. Such behavior suggests that the optimizer is struggling to find a continuous physical trajectory, instead trapped in a sequence of discrete, localized equilibria.
\item \textit{Parameter role-swapping and permutation symmetry:} We observe that parameters often “swap roles” between adjacent layers due to the inherent permutation symmetry of the QAOA ansatz. If a configuration $(\gamma_1, \gamma_2) \approx (a, b)$ yields a high-quality solution, we find that a perturbed permutation, such as $(\gamma_1, \gamma_2) \approx (b+\epsilon, a+\epsilon)$, frequently results in a nearly identical cost value. Standard gradient-free optimizers randomly “jump” between these symmetric equivalent points during the search process. This phenomenon creates erratic, non-smooth parameter sequences that lack temporal coherence and hinder convergence in deep circuits.
\item \textit{Landscape complexity across scaling regimes:} These pathologies are amplified as we scale the problem across different regimes:
\begin{itemize}
\item  When increasing the number of qubits while keeping the depth fixed (e.g., $p=2$), the density of the graph complicates the landscape, making the “jumps” more pronounced.
\item  In regimes with few qubits but very high depth ($p \gg 1$), the high-dimensional search space becomes unmanageable for traditional optimizers, leading to a breakdown in parameter stability.
\item  At the edge of our maximum simulation capacity, the accumulation of these uncontrolled sequences leads to a significant performance gap compared to functionally-defined schedules, necessitating a framework that enforces global structure over local independence.
\end{itemize}
\end{enumerate}

\paragraph{Contributions:}
The primary contribution of LOTUS (Layer-Ordered Temporally-Unified Schedules) lies in its fundamental restructuring of the QAOA variational landscape from a high-dimensional, chaotic search space into a structured, low-dimensional dynamical system. By replacing the standard layer-wise independent parameterization with a hybrid Fourier-Autoregressive (HFA) generator, LOTUS enforces temporal coherence and breaks permutation symmetry, effectively eliminating the $p!$ redundant minima that plague traditional optimization. Mathematically, LOTUS achieves a dimensionality collapse, reducing the optimization complexity to $O(1)$ relative to circuit depth $p$. This provides two transformative advantages:
\begin{enumerate}
\item It enables the training of deep circuits without the exponential degradation of classical optimizer performance;
\item It facilitates depth transferability, where continuous schedules optimized at low depths serve as near-optimal initializations for deeper circuits.
\end{enumerate}
Ultimately, LOTUS provides a robust pathway for scaling QAOA to the requirements of utility-scale quantum advantage, offering a superior balance between solution quality and computational efficiency.

\paragraph{Results:}
\begin{figure}[t]
\centering
\includegraphics[width=\linewidth]{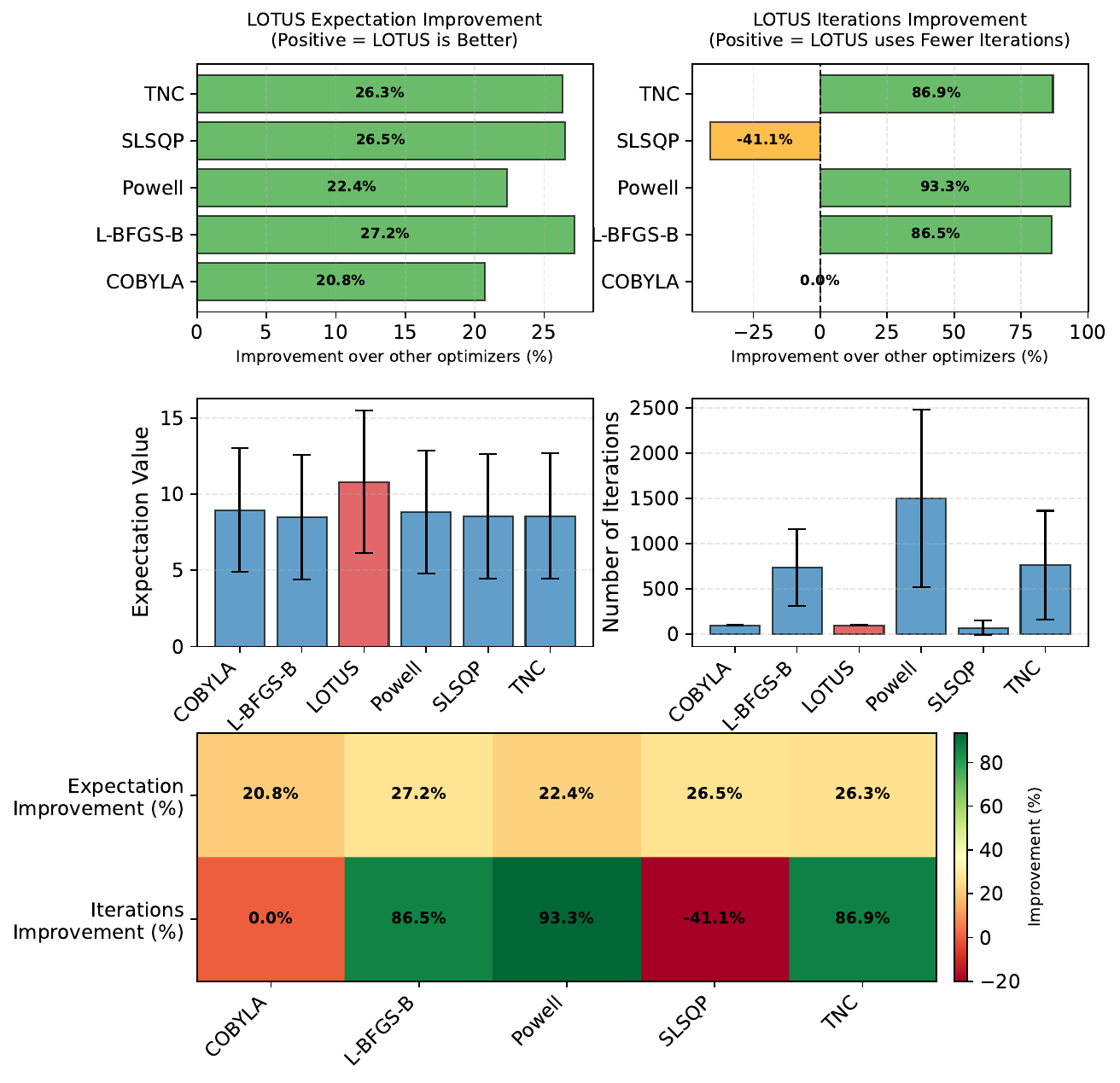}
\caption{Performance benchmarking of the LOTUS framework against traditional optimizers (L-BFGS-B, SLSQP, Powell, COBYLA, and TNC) for weighted MaxCut, showing significant percentage improvements in both expectation values (solution quality) and total iteration counts (computational efficiency)}
\label{fig:overview}
\end{figure}
Our analytical evaluation (Figure~\ref{fig:overview}) of the LOTUS framework reveals a decisive performance advantage over traditional optimization methods, characterized by a significant simultaneous improvement in both solution quality and computational efficiency. As illustrated in the expectation improvement analysis, LOTUS consistently yields higher expectation values across all tested baselines. It achieves a substantial $27.2\%$ improvement over L-BFGS-B and over $26\%$ improvement compared to TNC and SLSQP. This gain in solution quality does not come at a higher computational cost; rather, the “LOTUS Iterations Improvement” metric demonstrates that LOTUS is remarkably efficient, utilizing significantly fewer iterations to reach convergence in most cases. Specifically, LOTUS shows an iteration reduction of $93.3\%$ against Powell and over $86\%$ against TNC and L-BFGS-B. While most optimizers exhibit a trade-off between quality and speed, LOTUS breaks this Pareto bottleneck by leveraging its low-dimensional Hybrid Fourier-AR parameterization, which bypasses the chaotic, high-dimensional landscapes that typically hinder traditional gradient-free and gradient-based solvers. The cumulative result is a robust, scalable optimizer that transforms QAOA training from a high-overhead search into a streamlined functional refinement.

\section{Methods}\label{sec:method}
\subsection{Framework overview}
Standard parameterizations of QAOA treat the variational angles ${\gamma_l, \beta_l}_{l=1}^p$ as $2p$ independent degrees of freedom. This discrete independence introduces significant pathologies, including high-dimensional barren plateaus and problematic layer-permutation symmetries, where the objective function $\mathcal{L}$ remains invariant under the exchange of layer indices. We propose LOTUS (\textbf{L}ayer-\textbf{O}rdered \textbf{T}emporally-\textbf{U}nified \textbf{S}chedules), a framework that reframes QAOA parameterization as a constrained dynamical trajectory. In LOTUS, the variational parameters are generated by a functional mapping $f_l: \Phi \to \mathbb{R}^2$, such that $(\gamma_l, \beta_l) = f_l(\vec{\phi})$, where $\vec{\phi}$ is a low-dimensional vector of hyperparameters. This approach fundamentally restructures the optimization landscape in three ways:
\begin{enumerate}
\item By defining parameters as functions of the layer index $l$, we impose a structural ordering. The non-permutable nature of the basis functions (e.g., polynomials or sinusoids) ensures that each layer carries unique information, effectively eliminating the flat directions in the landscape associated with permutation symmetry.
\item The optimization is performed over the hyperparameter space $\Phi \subset \mathbb{R}^d$ where $d \ll 2p$. This dimensionality reduction mitigates the complexity of the classical optimization task, particularly as the circuit depth $p$ increases.
\item LOTUS enforces global coherence across the circuit depth. The “Temporally-Unified” nature of the schedules ensures that the parameters follow a smooth, physical trajectory—akin to a discretized adiabatic path—rather than chaotic, layer-wise independent values.
\end{enumerate}

\subsection{Framework details}
\paragraph{From discrete pathologies to continuous dynamics:}
In the standard ansatz, parameters $\Theta_{\text{std}} = { (\gamma_l, \beta_l) \mid l = 1, \dots, p } \in \mathbb{R}^{2p}$ lack inter-layer correlation. We instead treat $\gamma_l$ and $\beta_l$ as discrete samples from continuous functions of a normalized time coordinate $\tau \in [0, 1]$.

\paragraph{The hybrid Fourier-autoregressive (HFA) ansatz:}
The LOTUS framework implements a Hybrid Fourier-Autoregressive (HFA) mapping $\mathcal{M}: \mathbb{R}^{d} \to \mathbb{R}^{2p}$. We define the parameters for layer $l$ using a superposition of a global spectral trend and a local autoregressive perturbation. Let $x_l = \frac{l - 1/2}{p}$ be the normalized temporal coordinate; the parameters are defined as:
\begin{equation}
\begin{aligned}
\gamma_l(\vec{a}, \lambda_\gamma, \delta\gamma_0) &= \sum_{k=1}^{K} a_k \sin(k \pi x_l) + \delta\gamma_l \
\beta_l(\vec{b}, \lambda_\beta, \delta\beta_0) &= \sum_{k=1}^{K} b_k \cos(k \pi x_l) + \delta\beta_l
\end{aligned}
\end{equation}
where $K \ll p$ (modes) denotes the spectral bandwidth. This formulation comprises two distinct components:
\begin{enumerate}
\item The Fourier backbone: A truncated spectral expansion that captures the \textbf{global schedule} of the evolution. By restricting $k \leq K$, we impose a bandwidth limit that prohibits high-frequency “jitter,” enforcing smooth parameter trajectories.
\item The autoregressive (AR) moothing: To allow for local flexibility without reintroducing independence, the residuals $\delta\gamma_l$ and $\delta\beta_l$ are modeled as a First-Order Autoregressive process ($AR(1)$):
\begin{equation}\delta\gamma_{l} = \lambda_\gamma \cdot \delta\gamma_{l-1}, \quad \delta\beta_{l} = \lambda_\beta \cdot \delta\beta_{l-1}
\end{equation}
Here, $\lambda \in [0, 1]$ acts as a \textbf{stiffness} coefficient, ensuring that local deviations from the Fourier trend remain temporally correlated and decay or persist smoothly across layers.
\end{enumerate}

\paragraph{Optimization space:}
The optimization vector $\Theta$ is consolidated into a compact, structured tensor:
\begin{equation}
\Theta = \left[ \underbrace{a_1, \dots, a_K}{\text{Spectrum } \gamma}, \underbrace{b_1, \dots, b_K}{\text{Spectrum } \beta}, \underbrace{\lambda_\gamma, \delta\gamma_1}{\text{AR } \gamma}, \underbrace{\lambda\beta, \delta\beta_1}_{\text{AR } \beta} \right]
\end{equation}
This represents a mathematically rigorous separation of global spectral control and local adaptability, effectively solving the pathology of piecewise switching and providing a robust pathway for training deep quantum circuits.

\paragraph{Algorithms:} To bridge the theoretical framework of LOTUS with its practical implementation, we detail the underlying computational procedures in Algorithms \ref{alg:hfa_generator} and \ref{alg:hfa_optimization}, which formalize the transition from a low-dimensional hyperparameter space to a fully realized quantum circuit schedule.

\section{Theoretical analysis}
\subsection{Definitions}
We set up several definitions for the theoretical analysis of LOTUS, which are
\begin{definition}[Circuit Depth and Layer Indexing]
Let $p \in \mathbb{N}$ denote the QAOA circuit depth. We define the normalized temporal coordinate for each layer $l$ as
\begin{equation}
x_l := \frac{l - \frac{1}{2}}{p}, \qquad l = 1, \dots, p.
\end{equation}
\end{definition}
\begin{definition}[Standard QAOA Parameter Space]
The standard QAOA ansatz at depth $p$ is defined by the independent parameter set $\Theta_{\mathrm{std}}^{(p)} := \mathbb{R}^{2p}$, where $\theta := (\gamma_1, \dots, \gamma_p, \beta_1, \dots, \beta_p)$.
\end{definition}
\begin{definition}[LOTUS HFA Ansatz]
Fix a spectral bandwidth $K \ll p$. The Hybrid Fourier–AR (HFA) parameter space is defined as $\Theta_{\mathrm{HFA}}^{(K)} := \mathbb{R}^{2K+4}$, where $\phi := (a, b, \lambda_\gamma, \delta_{\gamma,0}, \lambda_\beta, \delta_{\beta,0})$. The mapping $\mathcal{M}p : \Theta{\mathrm{HFA}}^{(K)} \to \Theta_{\mathrm{std}}^{(p)}$ is given by:
\begin{equation}
\gamma_l = \sum_{k=1}^K a_k \sin(k\pi x_l) + \lambda_\gamma^{l-1} \delta_{\gamma,0}, \qquad \beta_l = \sum_{k=1}^K b_k \cos(k\pi x_l) + \lambda_\beta^{l-1} \delta_{\beta,0}.
\end{equation}
\end{definition}
\begin{definition}[QAOA Cost Function]
Let $\mathcal{C}(\phi; p) := \langle \psi_p(\phi) | H_C | \psi_p(\phi) \rangle$, where:
\begin{equation}
|\psi_p(\phi)\rangle = \prod_{l=1}^p e^{-i\beta_l H_B} e^{-i\gamma_l H_C} |+\rangle^{\otimes n}.
\end{equation}
\end{definition}

\subsection{Key theoretical Results}
\paragraph{Dimensionality and complexity:}
\begin{theorem}
For any depth $p$, the ratio of the HFA parameter space to the standard space vanishes as $p \to \infty$:
\begin{equation}
\lim_{p \to \infty} \frac{\dim(\Theta_{\mathrm{HFA}}^{(K)})}{\dim(\Theta_{\mathrm{std}}^{(p)})} = \lim_{p \to \infty} \frac{2K + 4}{2p} = 0.
\end{equation}
\end{theorem}
\textbf{\textit{Implication:}} For fixed $K$, the classical optimization problem over $\Theta_{\mathrm{HFA}}^{(K)}$ has $O(1)$ dimensional complexity with respect to circuit depth $p$. Consequently, the exponential degradation of optimizer performance with increasing depth, i.e., common in standard QAOA, is eliminated.

\paragraph{Enforced Lipschitz continuity and temporal coherence:}
\begin{theorem}[Layer-wise Lipschitz continuity]\label{thm:lipschitz}
Assume $|a_k| \le A, |b_k| \le A$, and $|\lambda_\gamma|, |\lambda_\beta| < 1$. Let ${\gamma_l}{l=1}^p$ be generated by the HFA ansatz under the above assumption, then there exist constants $C{\mathrm{spec}}, C_{\mathrm{AR}} > 0$, independent of $p$, such that:
\begin{equation}
|\gamma_{l+1} - \gamma_l| \le \frac{C_{\mathrm{spec}}}{p} + C_{\mathrm{AR}} |\lambda_\gamma|^{l-1}, \qquad \forall l.
\end{equation}
\end{theorem}
\textbf{\textit{Implication:}} Theorem~\ref{thm:lipschitz} implies $\lim_{p\to\infty} |\gamma_{l+1}-\gamma_l| = 0,$ uniformly away from the initial layers. Thus, the HFA ansatz forbids piecewise-constant switching and enforces a smooth dynamical trajectory.

\paragraph{Depth transfer and continuum invariance: }
\begin{theorem}[Depth Transfer Invariance]
Assume $|H_C|, |H_B| < \infty$. Then there exists a functional $\mathcal{F}[\phi]$ such that $\lim_{p \to \infty} \mathcal{C}(\phi; p) = \mathcal{F}[\phi]$, and for any $p, p' \gg 1$
\begin{equation}
\big| \mathcal{C}(\phi; p) - \mathcal{C}(\phi; p') \big| = O\left(\frac{1}{\min(p,p')}\right).
\end{equation}
\end{theorem}
\textbf{\textit{Implication:}} An optimizer solution $\phi^\star$ found at shallow depth $p$ remains $O(1/p)$-optimal at any deeper depth $p' \gg p$ via resampling of the same continuous schedules.

\section{Experimental settings}
\begin{figure}[t]
\centering
\includegraphics[width=\linewidth]{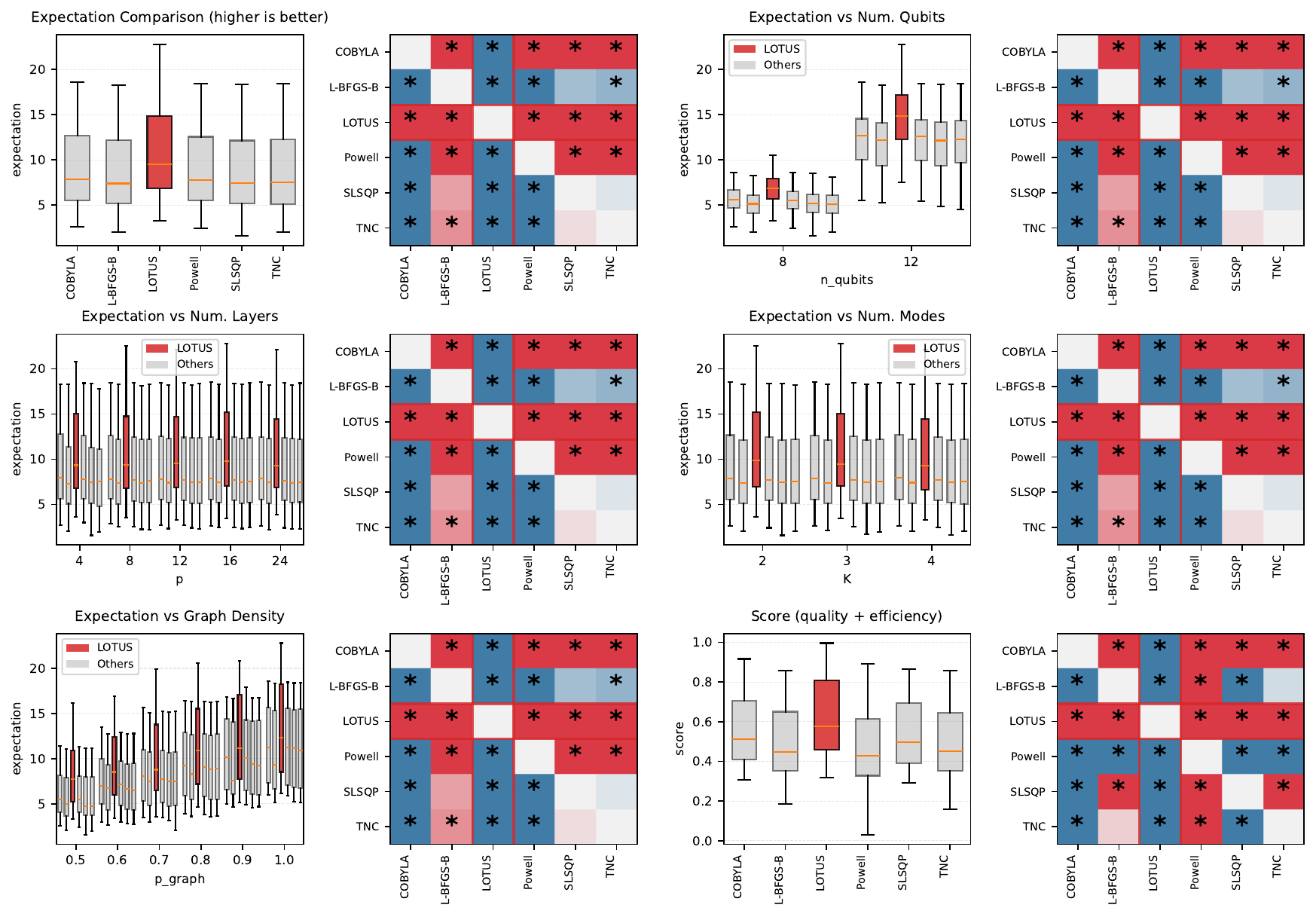}
\caption{Comparative results between LOTUS and other optimizers}
\label{fig:LOTUS}
\end{figure}
\paragraph{Setup:}
To evaluate the efficacy of the proposed LOTUS framework, we benchmark its performance against standard optimization strategies on the Weighted MaxCut problem. Our experiments focus on assessing trainability, parameter efficiency, and convergence stability in high-depth QAOA circuits. We generate synthetic problem instances using weighted Erdős-Rényi graphs $G(n, p_{\text{graph}})$. For our primary benchmark, we use the graph size with $n=8$ and $12$ nodes (qubits) with an edge probability from $p_{\text{graph}}=0.5$ to $1.0$. Edge weights are drawn uniformly from a continuous distribution $w_{ij} \sim U(0, 1)$, ensuring a non-trivial optimization landscape distinct from unweighted MaxCut. This setup creates connected graphs with dense connectivity patterns, serving as a robust testbed for evaluating optimizer performance in the presence of barren plateaus typical of dense Hamiltonians.

\paragraph{LOTUS configuration:} We employ QAOA with a circuit depth from $p=4$ to $p=24$ layers. The ansatz consists of alternating cost (Ising coupling) and mixer (Transverse field) unitaries.  Instead of optimizing the $2p=48$ raw variational parameters ($\vec{\gamma}, \vec{\beta}$) directly, we employ the LOTUS Hybrid Fourier-AR generator (Section~\ref{sec:method}). The generator constructs layer-wise parameters using a truncated Fourier series with $K=4$ modes, augmented by an autoregressive (AR) error correction mechanism. The trainable parameters $\vec{\theta}$ consist of:
\begin{itemize}
\item Fourier coefficients: Amplitude vectors $\mathbf{a}, \mathbf{b} \in \mathbb{R}^K$ for $\gamma$ and $\beta$ schedules.
\item AR parameters: Decay rates $\lambda_\gamma, \lambda_\beta$ and initial residuals $\delta_{\gamma,0}, \delta_{\beta,0}$.
\item Frequency weights: A weight vector $\mathbf{w} \in \mathbb{R}^K$ modulating the frequency contribution.
\end{itemize}
This results in a compressed parameter space of dimension $d_{\text{LOTUS}} = 3K + 4 = 16$, representing a $3\times$ compression factor compared to the standard parameterization ($d_{\text{std}} = 48$).

\paragraph{Baselines and optimization protocol:}
We benchmark LOTUS against five industry-standard gradient-free and gradient-based optimizers: COBYLA, L-BFGS-B, SLSQP, Powell, and TNC. All baselines optimize the full $2p$ parameter space directly. The cost function is the negative expectation value $-\langle H_C \rangle$, estimated via measurement sampling (MaxCUT objective). During the optimization loop, expectation values are estimated using 1024 shots per circuit evaluation. Final performance verification is conducted using 8192 shots to ensure high-precision readout of the best bitstring. Baseline optimizers are initialized with random parameters $\in [0, 2\pi]^{2p}$. The LOTUS optimizer employs a multi-start strategy with $n_{\text{restarts}}=5$, initializing Fourier coefficients with Gaussian noise ($\sigma=0.5$) and AR decay parameters in the stable region $[0.5, 0.95]$.

\paragraph{Evaluation metric:} To provide a comprehensive assessment that balances solution quality with convergence efficiency, we introduce a composite performance metric, denoted as the Score. This metric is defined as a weighted linear combination of the normalized expectation value and the inverted normalized iteration count:
\begin{equation}
\text{Score} = \alpha \cdot E_{\text{norm}} + (1 - \alpha) \cdot I_{\text{norm}}
\end{equation}
where we set $\alpha = 0.7$ to prioritize solution quality while retaining significant sensitivity to computational cost. The term $E_{\text{norm}}$ represents the solution quality, calculated via min-max normalization of the expectation value $\langle H_C \rangle$ within each seed group, such that $E_{\text{norm}} \in [0, 1]$ with 1 corresponding to the highest quality solution. Conversely, $I_{\text{norm}}$ quantifies algorithmic efficiency; it is computed by min-max normalizing the number of iterations and inverting the result, ensuring that fewer iterations map to a value closer to 1. This formulation ensures that the metric rewards optimizers that achieve superior energy minimization without incurring excessive computational overhead.

\section{Results}
\paragraph{LOTUS enhances the trainability of QAOA:}
Our results (Figure~\ref{fig:LOTUS}) demonstrate that LOTUS significantly outperforms standard optimizers in terms of solution quality. As shown in the \textbf{Expectation Comparison} analysis, LOTUS achieves a consistently higher median expectation value compared to the aggregated baseline optimizers (denoted as “Others”), with a tighter interquartile range indicating superior convergence stability. This performance advantage is further quantified in the \textbf{Score (quality + efficiency)} metric, where LOTUS exhibits a distinct advantage, with the majority of its distribution exceeding the median scores of competing methods like COBYLA and L-BFGS-B. The pairwise statistical significance matrices (heatmaps) corroborate this, displaying a dense concentration of asterisks in the LOTUS rows, confirming that the improvements over optimizers such as Powell, SLSQP, and TNC are statistically significant ($\alpha = 0.05$).

\paragraph{LOTUS consistency improves when scaling the number of qubits:}
LOTUS exhibits remarkable robustness as the system size scales. In the \textbf{Expectation vs Num. Qubits} analysis, while the performance of baseline optimizers shows a tendency to degrade or exhibit high variance as the number of qubits increases from $8$ to $12$, LOTUS maintains high expectation values. The box plots indicate that LOTUS effectively navigates the increasingly complex optimization landscape of larger Hilbert spaces (increasing number of qubits), preserving a median expectation well above the baselines even at larger qubit counts.

\paragraph{LOTUS consistency improves when scaling the number of layers:}
The trainability of QAOA typically suffers at higher depths due to the proliferation of local minima and barren plateaus; however, LOTUS mitigates these issues. \textbf{The Expectation vs Num. Layers} results show that as the circuit depth $p$ increases from $4$ to $24$, LOTUS sustains superior performance relative to standard optimizers. While the variance for baseline methods often expands significantly at deeper layers ($p=16, 24$), suggesting optimization difficulties, LOTUS retains a compact distribution and a high median expectation, effectively leveraging the additional parameters provided by increased depth.

\paragraph{LOTUS consistency across graph densities:}
LOTUS demonstrates broad applicability across varying problem structures. As illustrated in the \textbf{Expectation vs Graph Density} plot, LOTUS consistently yields higher expectation values across the full range of graph densities ($p_{\text{graph}}$ from $0.5$ to $1.0$). The separation between the LOTUS distribution and the “Others” category remains distinct regardless of the graph connectivity, indicating that the optimizer's efficacy is not dependent on specific graph topologies.

\paragraph{Effect of LOTUS configuration:} The number of modes does not impact the improvement of LOTUS; i.e., using two modes is efficient. An analysis of the \textbf{Expectation vs Num. Modes} results reveals that increasing the number of modes ($K$) beyond 2 yields negligible gains in solution quality. The distribution of expectation values for LOTUS remains statistically indistinguishable across $K=2, 3, \text{and } 4$. Since the computational overhead scales with the number of modes, this saturation in performance suggests that setting $K=2$ is the most efficient configuration. It strikes an optimal balance, providing the enhanced trainability of the LOTUS framework without incurring the unnecessary computational cost associated with higher-mode approximations. Examples of evaluation dashboard is illustrated in Figure~\ref{fig:8-24} and ~\ref{fig:12-24}.

\section{Discussions}
\paragraph{Breaking permutation symmetry:} In standard QAOA, the cost function is invariant under the permutation of layer indices, leading to $p!$ redundant minima and flat directions that hinder classical optimization. The LOTUS Fourier backbone resolves this by making the parameters $\gamma_l, \beta_l$ explicitly dependent on the normalized layer index $x_l = (l - 0.5)/p$. Since the basis functions $\sin(k \pi x_l)$ and $\cos(k \pi x_l)$ are distinct for each $l$, the parameters are strictly ordered in time. This structural ordering breaks the permutation symmetry, collapsing the redundant search space and forcing the optimizer to respect the causal trajectory of the quantum evolution.

\paragraph{Local flexibility via AR corrections:}
While the Fourier components capture the global adiabatic-like evolution necessary for algorithmic success, the AR term introduces controlled elastic deformation. This mechanism allows the ansatz to adapt to specific irregularities in the problem Hamiltonian—such as unique graph connectivity patterns—that a rigid spectral expansion might fail to capture. The AR process effectively allows the schedule to bend locally to minimize energy without breaking the global temporal coherence.

\paragraph{Scalability to large depth $p$:}
A primary bottleneck in QAOA is the linear growth of the parameter space with circuit depth. LOTUS achieves $O(1)$ complexity relative to $p$ by fixing the number of free parameters (e.g., $2K$ Fourier coefficients, 2 decay rates, and 2 initial residuals). This allows researchers to scale circuits to hundreds of layers ($p \geq 100$) without increasing the difficulty of the classical optimization loop, bypassing the dimensionality explosion that typically triggers barren plateaus.

\paragraph{Support for depth transfer:}
Because LOTUS defines parameters as continuous functions $\theta(\tau)$ rather than discrete lists, it naturally supports depth transferability. A solution optimized at depth $p$ yields continuous curves $\gamma(\tau)$ and $\beta(\tau)$ that can be resampled at any arbitrary depth $p' > p$. By evaluating these functions at new grid points $x'_l = (l+0.5)/p'$, one obtains a hot start initialization for deeper circuits that is already near-optimal, drastically reducing the total computational budget required for high-depth training.

\section{Conclusion}
Future work will focus on several key directions to expand the utility of this framework: (1) validating the effectiveness of the HFA ansatz on a broader class of NP-hard problems beyond MaxCut, such as Maximum Satisfiability (MaxSAT), Traveling Salesperson (TSP), and Quadratic Unconstrained Binary Optimization (QUBO); (2) investigating the synergy between LOTUS and QAOA variants, including Multi-Angle QAOA (MA-QAOA) and ADAPT-QAOA, to determine if spectral smoothing can further accelerate convergence in non-standard gate sets; (3) incorporating noise-model awareness into the autoregressive component, allowing the schedule to “bend” in response to specific hardware decoherence profiles while maintaining global temporal coherence; and (4) leveraging the depth transferability property of LOTUS, we will explore “hot-start” protocols for utility-scale quantum processors, where schedules optimized on classical simulators at low depth are resampled for execution on physical hardware at massive depths.

\paragraph{Declaration}
The methods and numerical evaluations presented in this work are solely the responsibility of the author. Computational support was provided by G.A.I.A QTech, LLC. The views and conclusions expressed do not represent Viettel.

\bibliography{output}
\bibliographystyle{unsrt}

\begin{appendices}
\section{Supplemental results}
\begin{figure}[h]
\centering
\includegraphics[width=0.8\linewidth]{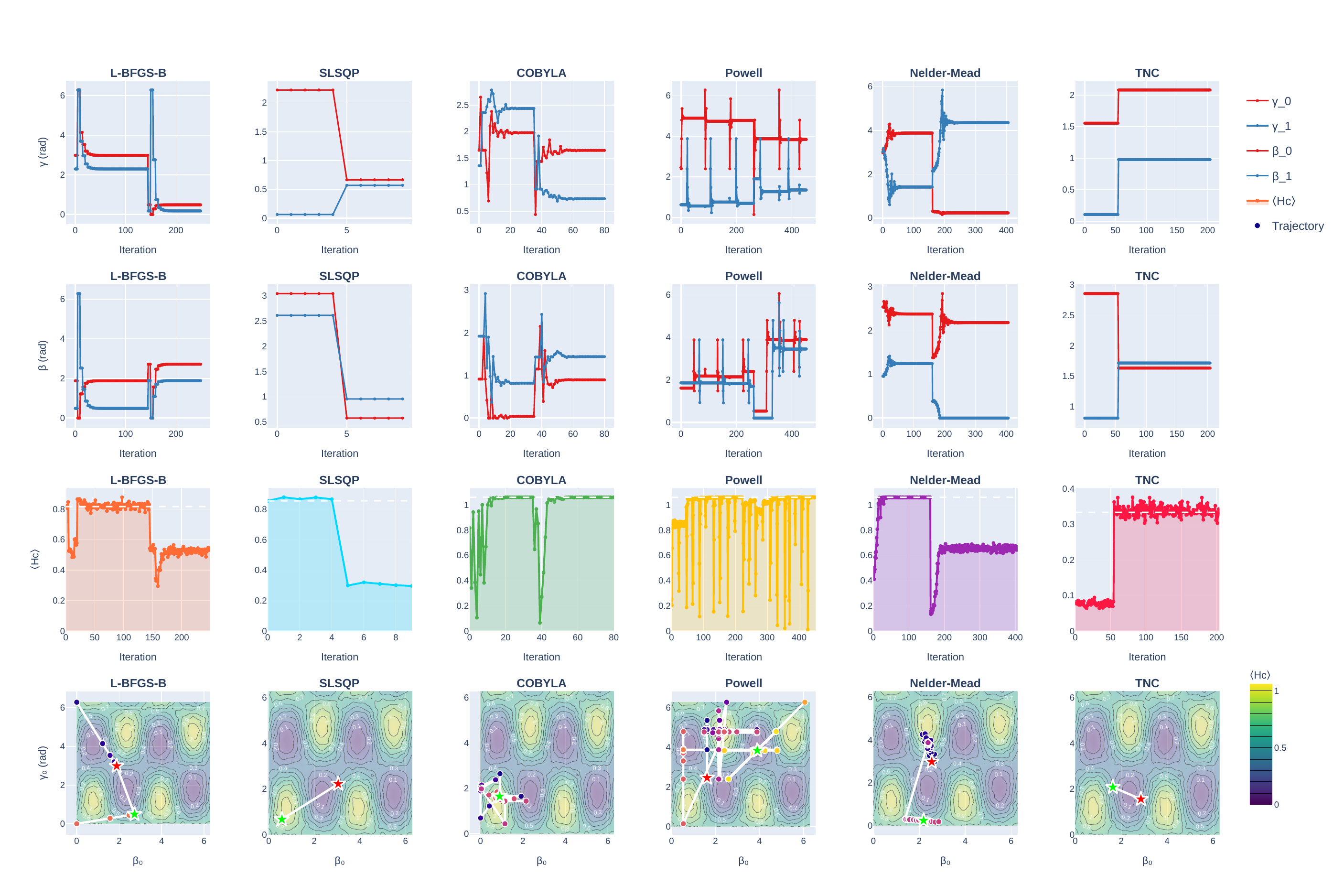}
\includegraphics[width=0.8\linewidth]{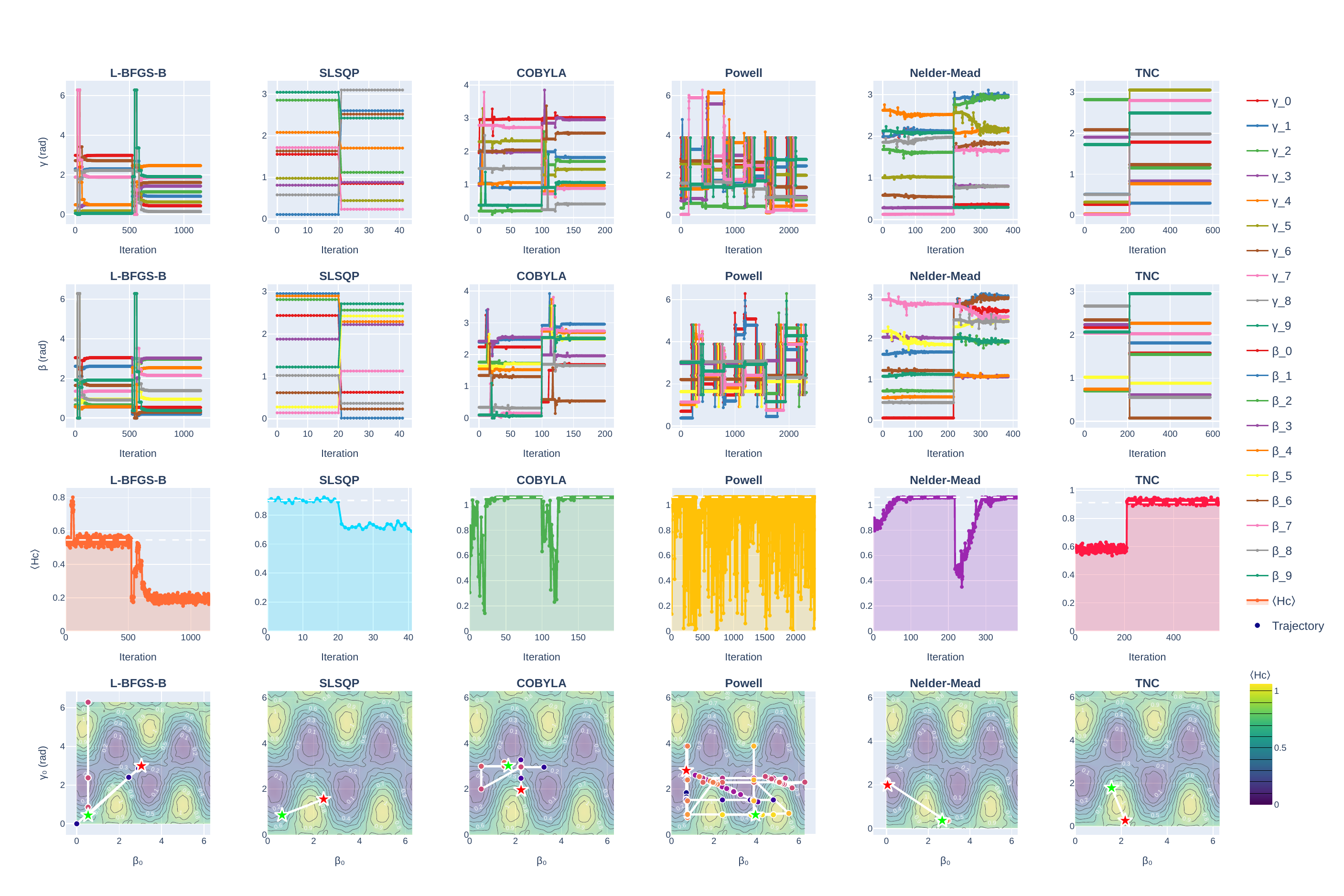}
\caption{Preliminary results of QAOA training with existing optimizers}
\label{fig:prelim_results}
\end{figure}
\begin{figure}[t]
\centering
\includegraphics[width= 0.75\linewidth]{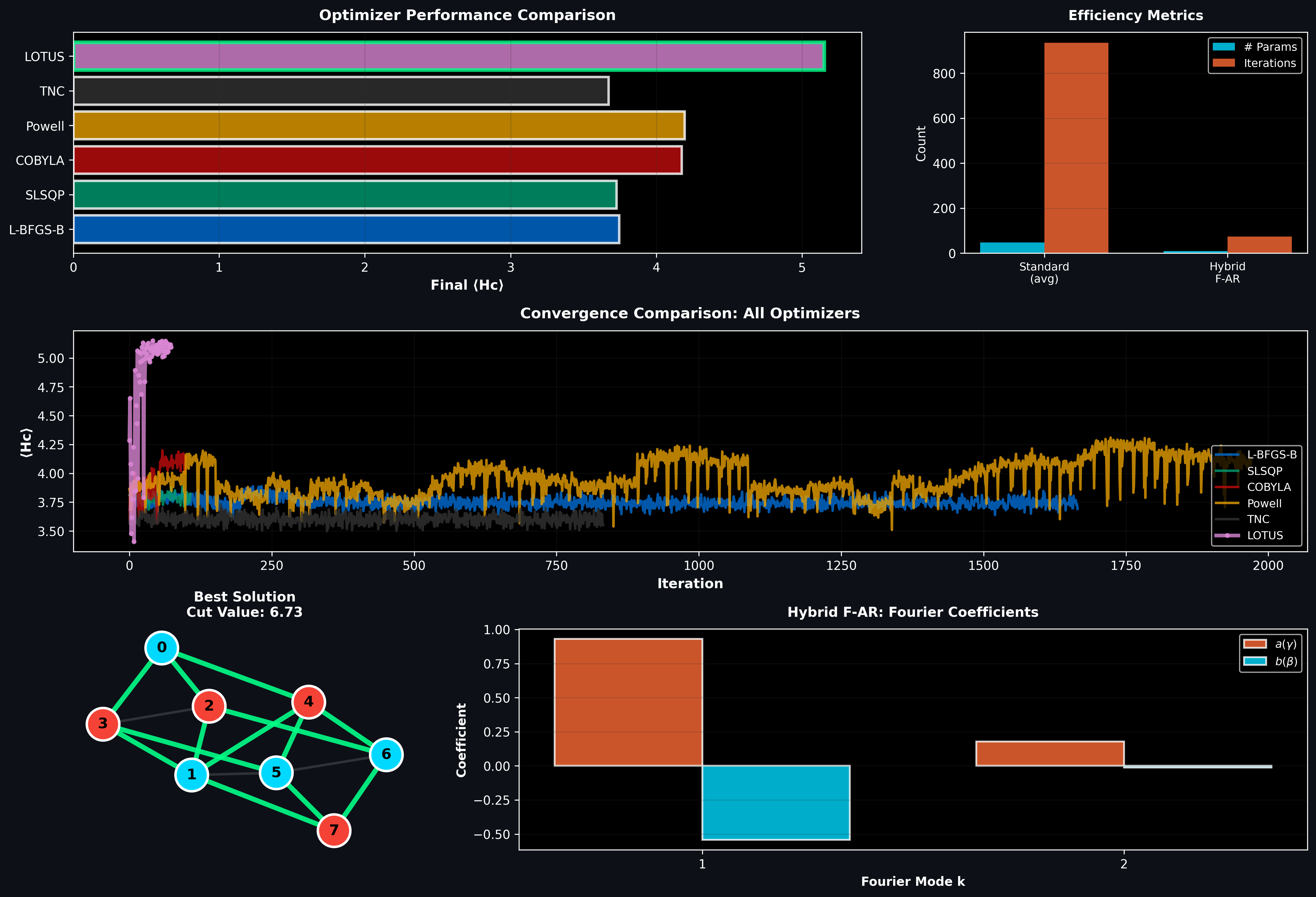}
\caption{Experimental results of LOTUS using $8$ qubits and $24$-layered QAOA}
\label{fig:8-24}
\end{figure}

\begin{figure}[h]
\centering
\includegraphics[width = 0.75\linewidth]{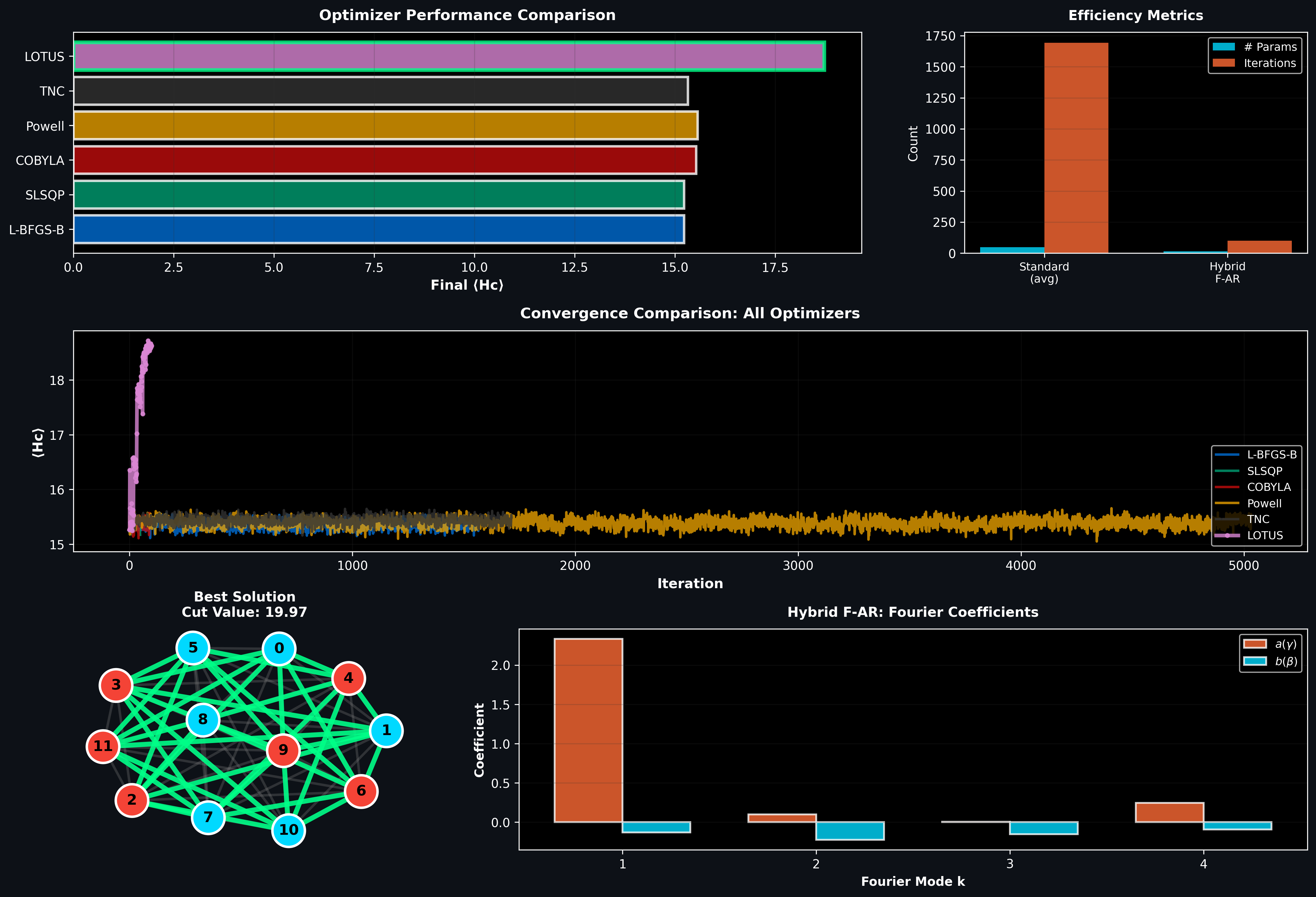}
\caption{Experimental results of LOTUS using $12$ qubits and $24$-layered QAOA}
\label{fig:12-24}
\end{figure}

\newpage
\section{Algorithms}
\begin{algorithm}[t]
\caption{Parameter Generation: $\mathtt{HybridFourierAR.generate}$}
\label{alg:hfa_generator}
\SetKwInput{KwInput}{Input}
\SetKwInput{KwOutput}{Output}
\SetKwComment{Comment}{$\triangleright$\ }{}

\KwInput{
Optimization vector $\Theta \in \mathbb{R}^{3K+4}$, 
Circuit depth $p$, Spectral bandwidth $K$.
}
\KwOutput{
QAOA parameter vectors $\vec{\gamma}, \vec{\beta} \in \mathbb{R}^p$.
}

\BlankLine

\Comment{1. Decomposition of $\Theta$ into Spectral and Autoregressive components}
$\mathbf{a} \leftarrow \Theta_{1 \dots K}, \quad \mathbf{b} \leftarrow \Theta_{K+1 \dots 2K}$ \Comment*[r]{Fourier Coefficients}
$\lambda_\gamma, \lambda_\beta \leftarrow \Theta_{2K+1, 2K+2}$ \Comment*[r]{AR Decay Rates}
$\delta_{\gamma,0}, \delta_{\beta,0} \leftarrow \Theta_{2K+3, 2K+4}$ \Comment*[r]{AR Initial Residuals}
$\mathbf{W} \leftarrow \Theta_{2K+5 \dots 3K+4}$ \Comment*[r]{Frequency Modulation Weights}

\BlankLine
\Comment{2. Compute Fourier Backbone and Autoregressive Residuals}
\For{$l \leftarrow 1$ \KwTo $p$}{
    $x_l \leftarrow (l - 1/2) / p$ \Comment*[r]{Normalized temporal coordinate}
    $F_{\gamma,l} \leftarrow \sum_{k=1}^{K} a_k \cdot W_k \cdot \sin(k \pi x_l) $\;
    $F_{\beta,l} \leftarrow \sum_{k=1}^{K} b_k \cdot W_k \cdot \cos(k \pi x_l) $\;

    \Comment{Iterative AR(1) Update}
    \eIf{$l = 1$}{
        $\delta\gamma_l \leftarrow \delta_{\gamma,0}, \quad \delta\beta_l \leftarrow \delta_{\beta,0}$\;
    }{
        $\delta\gamma_{l} \leftarrow \lambda_\gamma \cdot \delta\gamma_{l-1}$\;
        $\delta\beta_{l} \leftarrow \lambda_\beta \cdot \delta\beta_{l-1}$\;
    }
}

\BlankLine
\Comment{3. Synthesis and Physical Parameter Mapping}
\For{$l \leftarrow 1$ \KwTo $p$}{
    $\gamma_l \leftarrow (F_{\gamma,l} + \delta\gamma_l) \pmod{2\pi}$\;
    $\beta_l \leftarrow (F_{\beta,l} + \delta\beta_l) \pmod{2\pi}$\;
}

\Return $\vec{\gamma} = (\gamma_1, \dots, \gamma_p)$, $\vec{\beta} = (\beta_1, \dots, \beta_p)$
\end{algorithm}

\vspace{1em}

\begin{algorithm}[t]
\caption{LOTUS-QAOA Optimization Loop}
\label{alg:hfa_optimization}
\SetKwInput{KwInput}{Input}
\SetKwInput{KwOutput}{Output}
\SetKwProg{Fn}{Function}{:}{end}
\SetKwComment{Comment}{$\triangleright$\ }{}

\KwInput{
Problem Hamiltonian $H_C$, Mixer Hamiltonian $H_B$, 
LOTUS Generator $\mathtt{Gen}$, Classical Optimizer $\mathtt{CO}$.
}
\KwOutput{
Optimal hyperparameter vector $\Theta^*$, Best Expectation Value $E^*$.
}

\BlankLine
\Fn{$\mathcal{L}(\Theta)$}{
    $(\vec{\gamma}, \vec{\beta}) \leftarrow \mathtt{Gen.generate}(\Theta)$ \Comment*[r]{Map hyperparameters to angles}
    
    $|\psi(\Theta)\rangle \leftarrow \left( \prod_{l=1}^p e^{-i \beta_l H_B} e^{-i \gamma_l H_C} \right) |+\rangle^{\otimes n}$ \Comment*[r]{State Preparation}
    
    $E(\Theta) \leftarrow \langle \psi(\Theta) | H_C | \psi(\Theta) \rangle$ \Comment*[r]{Expectation via sampling}
    
    \Return $-E(\Theta)$\;
}

\BlankLine
\Comment{Global Optimization Procedure}
$\Theta_0 \sim \mathcal{N}(0, \sigma^2)$ \Comment*[r]{Initialize spectral coefficients}
$\Theta^* \leftarrow \mathtt{CO}.\text{minimize}(\mathcal{L}(\Theta), \Theta_0)$\;
$E^* \leftarrow -\mathcal{L}(\Theta^*)$\;

\Return $\Theta^*, E^*$
\end{algorithm}

\end{appendices}

\end{document}